\documentclass[prb,twocolumn,superscriptaddress,showpacs,
amsmath,amssymb]{revtex4}

\usepackage{graphicx} 
\usepackage{dcolumn} 

\def\b0{{\bf 0}}

\begin{document}

\title{Critical Casimir forces for the $O(N)$-models from functional renormalization }

\author{P. Jakubczyk}
\email{pjak@fuw.edu.pl}
\affiliation{Institute of Theoretical Physics, Faculty of Physics, University of Warsaw, 
 Ho\.za 69, 00-681 Warsaw, Poland}
\author{M. Napi\'{o}rkowski}
\affiliation{Institute of Theoretical Physics, Faculty of Physics, University of Warsaw, 
 Ho\.za 69, 00-681 Warsaw, Poland}
 
\date{\today}

\begin{abstract}
We consider the classical $O(N)$-symmetric models confined in a $d$-dimensional slab-like geometry and subject to periodic boundary conditions. Applying the one-particle-irreducible variant of functional 
renormalization group (RG) we compute the critical Casimir forces acting between the slab boundaries. The applied truncation of the exact functional RG flow equation retains interaction vertices of arbitrary order. We evaluate the critical Casimir 
amplitudes $\Delta_f(d,N)$ for 
continuously varying 
 dimensionality between two and three and $N = 1,2$. Our findings are in very good agreement with exact results for $d=2$ and $N=1$. For $d=3$ our results are closer to 
 Monte Carlo predictions than earlier field-theoretic RG calculations. Inclusion of the wave function renormalization and the corresponding anomalous dimension in the calculation has negligible impact 
 on the computed Casimir forces.    
 

\end{abstract}

\maketitle

\section{Introduction}
The concept of Casimir forces,\cite{Casimir_48, Mostepanienko_97, Krech_94, Kardar_99, Bordag_01, Gambassi_09, Brankov_00}  i.e. long-ranged effective interactions between macroscopic bodies immersed in
strongly fluctuating media, is nowadays recognized in plethora of systems ranging 
from biology to cosmology. On one hand their prominent importance stems from providing possibilities of testing theories of fundamental interactions,\cite{Masuda_09, Matsuda_11, Bezerra_2012, 
Klimchitskaya_12}   
on the other they are invoked to explain important mechanisms underlying physical phenomena in organic matter\cite{Bitbol_10, Machta_12} and as a control tool for nanodevices.\cite{Gambassi_09} 
The more conventional and well-studied 
(also experimentally) examples of Casimir forces in condensed matter include fluid mixtures\cite{Krech_97, Mukhopadhyay_00, Fukuto_95, Krech_99, Gambassi_09_2} and helium.\cite{Garcia_02, Dantchev_04, Ganshin_06, Maciolek_07, Hucht_07, 
Zandi_07, Hasenbusch_10}

The essence of the phenomenology underlying the Casimir forces is the presence of soft fluctuations occurring within a medium, whose spectrum of excitations becomes 
constrained by the boundary conditions imposed by the confining walls or by macroscopic objects immersed therein. As a result, the free energy acquires a contribution depending on the separation 
$L$ between the walls (or the objects) and the system finds it favorable to 
either increase or decrease this distance. A prominent characteristic of Casimir forces is the universality of its asymptotic behavior (including the amplitude) for $L\gg \Lambda_0^{-1}$ and 
$\xi\gg L$, where 
$\Lambda_0^{-1}$ 
is a system-specific microscopic lengthscale and $\xi$ denotes the (bulk) correlation length for the medium fluctuations. If the system features spontaneously broken continuous symmetry, 
one should additionally distinguish the transverse and longitudinal bulk correlation lengths. The system then displays a crossover between the Goldstone and critical regimes upon varying 
temperature. For a 
recent study of this crossover see Ref.~\onlinecite{Dohm_13}. On the other hand, the amplitude of the force (even its sign) does depend on the boundary conditions. The orthodox QED Casimir force\cite{Casimir_48} is a 
pure effect of quantum 
vacuum fluctuations. Its theoretical description involves a free field theory, where interactions with matter are taken into account via  boundary conditions and the expectation value of the fluctuating electromagnetic 
field is zero. The situation is quite distinct in these respects in many condensed matter systems. Here a field-theoretic description of the problem often requires a treatment of interaction terms in the 
effective action. For certain cases, the boundary terms may also lead to nonzero expectation values of the (nonuniform) order parameter, which gives rise to effective interactions between the confining 
walls (or other macroscopic objects) also in the absence of fluctuations. 

In this work we analyze the critical Casimir effect employing a class of standard models to describe different aspects of critical behavior, namely the $d$-dimensional $O(N)$-symmetric $\phi^4$-type models. 
The system is confined to be finite in one direction (so that its extension is $L$) and we consider the periodic boundary conditions. Despite not being of the highest relevance for quantitative comparison with 
experimental data, this choice of boundary conditions does not yield Casimir forces at mean-field level, so that the effective interaction is a pure fluctuation effect (preserving the analogy to the QED setup). In addition, 
these boundary conditions yield a zero-mode in the free propagator, which leads to substantial complications in a conventional RG treatment, where the $\epsilon$-expansion\cite{Krech_92} becomes 
ill-defined beyond two-loop order.\cite{Diehl_06, Gruneberg_08}    We study the 
system within the one-particle-irreducible (Wetterich) framework of functional RG.\cite{Wetterich_93, Berges_02} The exact flow equation governing the flow of the generating functional for irreducible vertices is treated within an 
approximation retaining the flow of interaction couplings of arbitrary order, but neglecting the renormalization of the momentum dependencies in the propagators. This amounts to the leading order of an 
approximation strategy known as derivative expansion,\cite{Berges_02} which was successfully applied \textsl{inter allia} for accurate calculations of the $O(N)$ bulk critical behavior\cite{Berges_02, 
Canet_03, Ballhausen_04}
(including the Kosterlitz-Thouless case\cite{Gersdorff_01}). 
This approach is not controlled by any small parameter, but may be considered as a leading order of a systematic procedure to compute critical properties. It can be applied for different $d$ and $N$. In 
particular, we scan the dependence of the critical Casimir force amplitude $\Delta_f(d,N)$ as a function of dimensionality $d$, varying $d$ between two and three at fixed $N\in\{1,2\}$. 
Our findings compare very well to 
the exact value\cite{Cardy_86} for $d=2$ and $N=1$ and reasonably well with numerical Monte Carlo (MC) simulations\cite{Dantchev_04_2, Vasilyev_09} in $d=3$ (both for $N=1$ and $N=2$).  

The paper is organized as follows: In Sec.~II we present the model and the RG setup. In Sec.~III we consider the bulk critical behavior upon varying $d$ between two and three at fixed number of 
order-parameter components $N\in \{1,2\}$. In particular, within the applied approximation, we compute the correlation length in the vicinity of the critical point and extract the values of the critical 
exponent $\nu (d,N)$, describing the 
divergence of the correlation length $\xi$ at criticality. In the subsequent Sec.~IV we present the computation of the critical Casimir forces in the cases $N=1$ and $N=2$, continuously varying $d$. 
In Sec.~IVC we gauge the accuracy of the applied exact RG truncation by performing a refined calculation including the wave function renormalization in the case $N=1$, $d=2$. We show that such a 
refinement has a negligible impact on the obtained Casimir forces. We summarize the work in Sec.~V.   

\section{Model and RG setup}
The Landau-type model considered here is a paradigm of the theory of phase transitions and critical phenomena. In particular, in Refs.~\onlinecite{Krech_92, Diehl_06, Gruneberg_08}
it was employed to study the thermal Casimir forces. The 
fluctuating $N$-component real order-parameter field $\vec{\chi}(\vec{x})$ is confined in a $d$-dimensional box of volume $V=D^{d-1}\times L$, where $D$ is much larger than all the other lengthscales 
present in the system. This imposes an idealization as compared to the systems realized in experiments and simulations, where the aspect ratio $L/D$ is always finite. For a discussion of the aspect ratio 
dependencies of Casimir forces see Refs.~\onlinecite{Dohm_09, Hucht_11}.

We impose periodic boundary conditions in all the directions and the effective Hamiltonian takes the standard form 
\begin{equation}
 \mathcal{S}[\vec{\chi}]=\int_V d^dx[U_0(\rho)+\frac{1}{2}Z_0(\nabla\vec{\chi})^2]\;, 
 \label{LGW}
\end{equation}
where we introduced $\rho=\frac{1}{2}(\vec{\chi})^2$, and the effective potential 
\begin{equation}
 U_0(\rho)=\frac{v}{2}\rho^2+\delta \rho
 \label{Initial_U}
\end{equation}
is quadratic in $\rho$. The model involves the interaction coupling $v$, the mass-like parameter $\delta$, the constant $Z_0$, and 
 is additionally supplemented with a lower cutoff in real space at $\Lambda_0^{-1}$, which we treat on equal footing with the other parameters. The parameter $\delta$ may be tuned by varying temperature, 
 pressure or some other physical quantity. At mean-field level its critical value is $\delta_c^{MF}=0$. On the other hand, accounting for fluctuation effects favors the disordered phase and reduces this value, 
 so that $\delta_c<0$.
 
The present approach\cite{Berges_02, fRG_reviews} relies on an exact RG equation governing the flow of the scale-dependent effective action $\Gamma_\Lambda [\vec{\phi}]$, where the field $\vec{\phi}$
is the expectation value of $\vec{\chi}(\vec{x})$.\cite{Berges_02}
Upon decreasing the momentum cutoff scale $\Lambda$  
from $\Lambda_0$ to 
zero, the functional $\Gamma_\Lambda [\phi]$ interpolates between the initial action $\mathcal{S}[\vec{\chi}=\vec{\phi}]$ at $\Lambda=\Lambda_0$ and the full effective action, i.e. the free energy 
$\Gamma_0 [\vec{\phi}]$ at $\Lambda=0$. 
The quantity $\Gamma_\Lambda [\vec{\phi}]$ may be understood\cite{Berges_02} as the free energy obtained after including the fluctuation modes with momentum $q\in [\Lambda, \Lambda_0]$. At $\Lambda=\Lambda_0$ no 
fluctuations are included and $\Gamma_\Lambda [\vec{\phi}]$ is just the bare effective action. On the other hand, in the limit $\Lambda \to 0$ all the fluctuation modes are included. 
The flow of $\Gamma_\Lambda [\vec{\phi}]$ upon reducing the cutoff 
scale $\Lambda$ is governed by the flow equation\cite{Wetterich_93}
\begin{equation}
\partial_\Lambda \Gamma_\Lambda[\vec{\phi}]=\frac{1}{2}\tilde{\mathrm{Tr}}\frac{\partial_\Lambda R_\Lambda (q^2)}{\Gamma^{(2)}_\Lambda[\vec{\phi}]+R_\Lambda (q^2)}  \;, 
\label{Wetterich_eq}
\end{equation}
where $R_\Lambda(q^2)$ is a momentum cutoff function added to the inverse propagator to regularize the infrared behavior, and 
$\Gamma^{(2)}_\Lambda[\vec{\phi}]=\delta^2\Gamma_{\Lambda}[\vec{\phi}]/(\delta \phi^a_{\vec{q}}\delta \phi^a_{-\vec{q}})$, with $a\in\{1,...N\}$ 
(no off-diagonal components of the inverse propagator occur in the present problem). The trace $\tilde{\mathrm{Tr}}$ sums over field components and momenta: 
$\tilde{\rm{Tr}}=\sum_{a}\frac{1}{L}\sum_{q_1}\int\frac{d^{d-1}q}{(2\pi)^{d-1}}$ in the limit $D\to\infty$ and with $q_1=\frac{2\pi n}{L}$, $n\in \mathbb{Z}$ . 
Integration of Eq.~(\ref{Wetterich_eq}) with the initial condition at $\Lambda_0$ 
specified by Eqs.~(\ref{LGW}) and (\ref{Initial_U}) yields the exact solution to the model. Herein we apply an approximation strategy known as derivative expansion,\cite{Berges_02} 
where the symmetry-allowed terms in $\Gamma_\Lambda [\vec{\phi}]$ are classified according to increasing number of derivatives and we keep only 
the leading (zeroth) 
order term in this expansion. This truncation level (the so-called 
local potential approximation (LPA)) amounts to imposing the following ansatz: 
\begin{equation}
\Gamma_\Lambda [\vec{\phi}] =  \int_V d^dx[U_\Lambda(\rho)+\frac{1}{2}Z_0(\nabla\vec{\phi})^2]\;, 
\label{LPA_ansatz}
\end{equation}
where we put $\rho=\frac{1}{2}(\vec{\phi})^2$.
The present approximation therefore neglects renormalization of all the momentum dependencies in the propagator, but does not truncate interaction terms in the local effective potential 
$U_\Lambda(\rho)$ at any order. In Sec.~IVC we perform an additional calculation capturing the wave function renormalization for $d=2$ and $N=1$. Our results indicate that the effect of such a truncation 
refinement on the Casimir forces is negligible even though it significantly improves the quality of the obtained bulk critical exponents. 

With the ansatz (\ref{LPA_ansatz}) the effective action flow is represented by the flow of the scale-dependent effective potential
\begin{equation}
\partial_\Lambda U_\Lambda(\rho) = \frac{1}{2}\mathrm{Tr}\left\{\partial_\Lambda R_\Lambda (q^2)\left[\frac{1}{M_R(\rho, q^2)}+\frac{N-1}{M_T(\rho, q^2)}\right]\right\}\;,
\label{eff_pot_flow}
\end{equation} 
where 
\begin{eqnarray}
M_R(\rho, q^2) = Z_0q^2+R_\Lambda (q^2)+U_\Lambda '(\rho)+2\rho U_\Lambda ''(\rho)\;,  \nonumber\\
M_T(\rho, q^2) = Z_0q^2+R_\Lambda (q^2)+U_\Lambda '(\rho)\;, \;\;\;\;\;\;\;\;\;\;\;\;\;\;\;\;\;\;
\label{Masses}
\end{eqnarray}
and $\rm{Tr}=\frac{1}{L}\sum_{q_1}\int\frac{d^{d-1}q}{(2\pi)^{d-1}}$.
Since we neglect the renormalization of $Z_0$, Eq.~(\ref{eff_pot_flow}) is closed and our subsequent analysis is based on its numerical integration. Any practical calculation requires specifying the 
cutoff function $R_\Lambda (q^2)$, which must respect a number of conditions.\cite{Berges_02} We make the following choice,\cite{Litim_01} known as the Litim cutoff
\begin{equation}
R_\Lambda (q^2)=Z_0(\Lambda^2-q^2)\theta (\Lambda^2-q^2)\;,
\label{Litim_cutoff}
\end{equation}
for which the trace in Eq.~(\ref{eff_pot_flow}) can be straightforwardly performed in the limit $L\to\infty$. 

The free energy is identified as
\begin{equation}
 F(\phi,L) = k_B T \Gamma _{\Lambda\to 0} [\phi=\phi_{unif}] = k_B T V U_{\Lambda\to 0}(\rho)\;,
\end{equation}
where the functional is evaluated at a uniform field configuration. Its dependence on $L$ arises due to the trace involving the $L$-dependent momenta allowed by the boundary conditions. The Casimir 
forces, in turn, result from the $L$-dependent free energy. Since our study concerns the long-ranged interactions close to a second-order phase transition, 
we first need to resolve the critical behavior, which is done in the next section.

\section{Critical behavior}
In order to efficiently tune the system to its bulk criticality, we consider the limit $L\to\infty$ and perform the standard variable transformation
\begin{eqnarray}
 \tilde{\rho}=Z_0\Lambda^{2-d}\rho \;\;\;\;\;\;\;\;\;\; \nonumber \\
 u_\Lambda^{\infty}(\tilde{\rho})=\Lambda^{-d}U_\Lambda(\rho)\;, 
\end{eqnarray}
where the superscript $\infty$ indicates that we deal with an infinite system. 
This brings Eq.~(\ref{eff_pot_flow}) to a scale-invariant form 
\begin{eqnarray}
\partial_s u_\Lambda^\infty (\tilde{\rho}) = d u_\Lambda^\infty (\tilde{\rho}) + (2-d)\tilde{\rho} {u_\Lambda^{\infty}}'(\tilde{\rho})  - \nonumber \\
A_{d}\left[\frac{1}{m_R(\tilde{\rho})} + \frac{N-1}{m_T(\tilde{\rho})}\right]\;, \;\;\;
\label{u_flow}
\end{eqnarray} 
where 
\begin{eqnarray}
 m_R(\tilde{\rho}) = 1 + {u_\Lambda^{\infty}}'(\tilde{\rho}) + 2\tilde{\rho}{u_\Lambda^{\infty}}''(\tilde{\rho}) \;,  \nonumber \\
 m_T(\tilde{\rho}) = 1 + {u_\Lambda^{\infty}}'(\tilde{\rho})\;. \;\;\;\;\;\;\;\;\;\; \;\;\;\;\;\;\;\;\;\;
\end{eqnarray}
The primes denote differentiation with respect to the argument. We introduced $s=-\ln(\Lambda/\Lambda_0)$ and $A_d=\frac{S^{(d-1)}}{(2\pi)^{d}d}$, where $S^{d}$ is the surface area of a $d$-dimensional unit sphere. 

We numerically integrate the discretized version of Eq.~(\ref{u_flow}) taking Eq.~(\ref{Initial_U}) as the initial condition. This is done by considering a grid in the $\tilde{\rho}$-space. The discrete 
$\tilde{\rho}$-derivatives were computed numerically using a 5-point routine at each RG step. For accuracy control we varied the grid size between $M=60$ and $M=200$ points. 
This discretization procedure therefore maps the partial differential Eq.~(\ref{u_flow}) onto a set of $M$ coupled nonlinear ordinary differential equations. This system can be handled relying on the 
standard available numerical stepping procedures including the simplest Euler-type routines.

For all of the analysis of the present section we put $Z_0=1$ and $\Lambda_0=1$. At fixed $d$ and the 
(initial) interaction coupling $v$ in 
Eq.~(\ref{Initial_U}), we scan the solution to Eq.~(\ref{u_flow}) for $\Lambda\to 0$ upon varying the bare mass parameter $\delta$ in Eq.~(\ref{Initial_U}). We identify two regimes in the space  spanned by 
$(\delta, v)$, corresponding to the symmetric and the symmetry-broken phases. The line of critical points (in the $(\delta, v)$ space) and the corresponding fixed points 
are found by the dichotomy procedure illustrated in Fig.~1.
\begin{figure}[h]
\includegraphics[width=8.5cm]{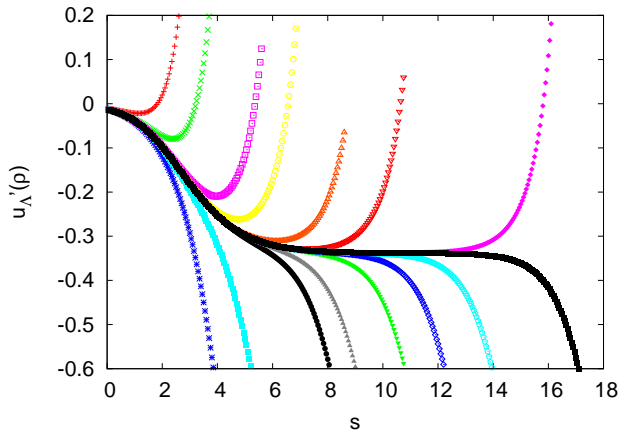}
\caption{Illustration of the dichotomy procedure employed to identify the critical point $\delta=\delta_c$. The present exemplary plot corresponds to $v=0.1$, $N=1$, and $d=2.6$. 
For $\delta>\delta_c$ the plotted flowing coupling 
${u_\Lambda^{\infty}}'(\tilde{\rho}=0)$ attains positive values at a finite scale, indicating flow to the symmetric phase. For $\delta<\delta_c$ the coupling ${u_\Lambda^{\infty}}'(\tilde{\rho}=0)$ 
decreases upon reducing the scale $\Lambda$ as the system flows to the symmetry-broken phase. Fixed-point behavior represented by the plateau corresponds to the critical value $\delta=\delta_c\approx -0.01253855209$. }
\end{figure}  
Having identified the critical point, we vary $\delta$ so that the system remains in the symmetric phase close to $\delta_c$ and scan the behavior of the (non-rescaled) effective potential at its minimum 
in the limit $\Lambda\to 0$. By computing the curvature of the effective potential as function of $\phi$ at $\phi=0$ for different values of $\delta>\delta_c$ in the vicinity of the critical point, we extract the correlation length 
$\xi(\delta)$. The obtained behavior very well fits the expected power-law 
\begin{equation}
 \xi\sim (\delta-\delta_c)^{-\nu}\;.
\end{equation}
We analyze the dependence of the exponent $\nu(d,N)$ focusing on $N\in\{1,2\}$ and varying $d$ between two and three. 
The obtained dependencies are shown in Figs 2 and 3 for $N=1$ and $N=2$, respectively. 
\begin{figure}[h]
\includegraphics[width=8.5cm]{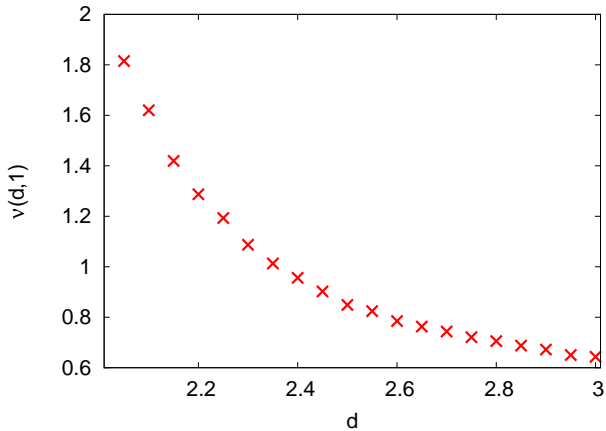}
\caption{The correlation length exponent $\nu(d,1)$ plotted as function of system dimensionality $d$ at $N=1$. The value $\nu(3,1)\approx 0.64$ agrees  well with the best available estimates. 
For $d$ approaching 2 the obtained result substantially deviates from the exact value $\nu_{exact}(2,1)=1$. See the main text for a discussion of this discrepancy.}
\end{figure} 
\begin{figure}[h]
\includegraphics[width=8.5cm]{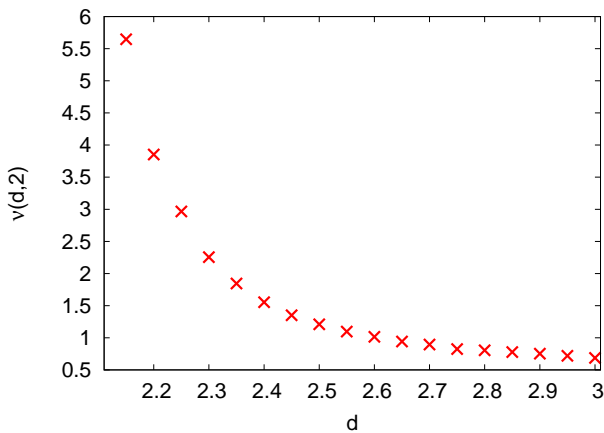}
\caption{The correlation length exponent $\nu(d,2)$ plotted as function of system dimensionality $d$ at $N=2$. The value $\nu(3,2)\approx 0.68$ compares  well to the best available estimates. 
We observe a rapid growth of $\nu (d,2)$ upon reducing $d$ toward 2. This tendency is in line with the expectation that the $XY$ model in $d=2$ features an essential singularity of the correlation length 
when approaching criticality from the symmetric phase.}
\end{figure} 
In both cases 
the values of $\nu$ compare well to the best available estimates\cite{Pelissetto_02} in $d=3$. We obtain $\nu(3, 1)\approx 0.64$ and $\nu(3, 2)\approx 0.68$ 
(the best estimates according to Ref.~\onlinecite{Pelissetto_02} are $\nu(3, 1)\approx 0.63$ and $\nu(3, 2)\approx 0.67$ up to two digits).  
 On the other hand, our results for $N=1$ and $d$ approaching 2
substantially deviate from the exact value $\nu_{exact}(2,1)=1$. 
This discrepancy arises because the present exact RG truncation neglects renormalization of the momentum dependencies in 
the propagator. Indeed, the singularity of the propagator renormalization at criticality is quantified by the value of the anomalous dimension $\eta$, which is relatively small in $d=3$ 
($\eta_3\approx  0.036$ for $N=1$), but sizable ($\eta_2=\frac{1}{4}$ for $N=1$) for $d=2$. The present approximation, Eq.~(\ref{LPA_ansatz}), obviously yields $\eta=0$. 
This also perturbs the very structure of Eq.~(\ref{u_flow}) at $d=2$, where the vanishing factor $(2-d)$ should be replaced by $(2-d-\eta)$. For this reason we did not achieve true 
fixed point behavior at $d=2$ and do not present the value of $\nu$ precisely for $d=2$ (see Fig.2). This however does not create any obstacle in computing the Casimir forces in $d=2$ and $N=1$ close to the critical 
point in the symmetric phase (see the next section). We readdress the case $d=2$, $N=1$ with a refined approximation capturing $\eta>0$ in Sec.~IVC.

For the same reason we are not able 
to resolve the critical behavior of 
the Kosterlitz-Thouless universality class ($d=2$, $N=2$), where nonzero $\eta$ is crucial to capture the physics even at a qualitative level. Interestingly, for $N=2$, we observe a rapid increase of 
$\nu (d,2)$ upon reducing $d$ towards 
2. This may indicate the divergence of $\nu (d,2)$ and the onset of the Kosterlitz-Thouless physics, where the correlation length exhibits an essential singularity in the symmetric phase. We also note, that the 
computed dependence $\nu(d,2)$ nicely fits a power law $\nu (d,2)\sim(d-2)^{-A}$ in the interval $d\in [2.15,2.4]$. We obtain $A\approx 1.31$. We also note, that the values of $\nu (d,N)$ in low $d$ 
significantly 
improve upon including $\eta$, i.e. going to higher order in derivative expansion. For a calculation resolving the $d$-dependence for $N=1$ at the next-to leading order in derivative expansion see 
Ref.~\onlinecite{Ballhausen_04}. 

One may at this point ask, to which extent the inaccuracy in resolving the bulk critical behavior in $d\approx 2$ will affect the results for the Casimir forces, the amplitudes in particular. 
At the outset one possibility is that the computed properties simply come out inaccurate. Many critical quantities are however insensitive to the value of $\eta$. In the present context, for example, 
it is known that the critical Casimir forces decay asymptotically as $L^{-d}$ irrespective of the value of $\eta$. The asymptotic behavior of the correlation function (involving $\eta$) therefore does not 
directly influence the power law governing the decay of the Casimir force. Another example one may invoke here is the universal exponent $\psi$ governing the shape of the finite-temperature transition 
line in 
quantum critical systems.\cite{Millis_93, Jakubczyk_08, Strack_09} Despite being located in the portion of the phase diagram which is dominated by non-Gaussian thermal fluctuations, $\psi$ is fully 
determined by the system dimensionality $d$ and the dynamical exponent $z$. The precise values of $\eta$ or $\nu$ are therefore not relevant to it. The results of the subsequent section suggest that the 
Casimir 
amplitudes also belong to this category. Indeed, our calculation yields a very accurate estimate of the critical Casimir amplitude in $d=2$, where bulk critical exponents are off the correct values and 
overall the approximation is less reliable than in the case $d\approx 3$. 

To support the above statement, and the conclusions of our computations, in Sec.~IVC we perform an additional calculation including the flow of 
$Z_0$ in $d=2$, $N=1$. This cures the drawbacks of the above analysis for $L\to\infty$, leads to significant improvement of the bulk critical exponents, but, as turns out, has negligible impact on the 
obtained Casimir forces.

\section{Casimir forces}
Having resolved the critical behavior for the infinite system, we now consider finite $L$ and proceed by computing the Casimir forces. The system state is chosen in the symmetric phase ($\delta>\delta_c$), but close enough to the 
critical point, so that the correlation length $\xi\gtrsim 10^5 \Lambda_0^{-1}$. The $L$-dependent contribution to the free energy resulting in the appearance of the Casimir force is 
identified as 
\begin{equation}
 \sigma(L)=\lim_{\infty}\frac{F-fV}{D^{d-1}}\;,
\end{equation}
where 
\begin{equation}
f=\lim_\infty\frac{F}{V}\;.
\end{equation}
This leads to 
\begin{equation}
 \sigma(L)=k_BTL\left[U_{\Lambda\to 0}^L(0)-U_{\Lambda\to 0}^{\infty}(0)\right]\;.
 \label{sigma}
\end{equation}
 The additional superscript in the effective potential $U_\Lambda^L (\rho)$ indicates the finite size dependence that we set out to compute now. The potential $U$ is evaluated at $\rho=0$, 
 because, as indicated above, we perform the computations in the symmetric phase. Utilizing Eq.~(\ref{eff_pot_flow}) with the cutoff given in Eq.~(\ref{Litim_cutoff}), we have
\begin{equation}
\partial_\Lambda U_{\Lambda}^{\infty}(\rho)=A_{d}Z_0\Lambda^{d+1}\left[\frac{1}{M_R^{\infty}(\rho)}+\frac{N-1}{M_T^{\infty}(\rho)}\right],
\label{U_Casi1}
\end{equation}
and 
\begin{eqnarray}
\partial_\Lambda U_{\Lambda}^{L}(\rho)=A_{d-1}L^{-1}Z_0\Lambda^d \left[\frac{1}{M_R^{L}(\rho)}+\frac{N-1}{M_T^{L}(\rho)}\right]\times \nonumber\\
\times\left[1+2\sum_{n=1}^{n_{max}}\left(1-\left(\frac{2\pi n}{\Lambda L}\right)^2\right)^{\frac{d-1}{2}}\right] \;,\;\;\;\;\;
\label{U_Casi2}
\end{eqnarray}
where $n_{max}=\frac{\Lambda L}{2\pi}$ and the quantities $M_{...}^{...}(\rho, q^2)$ are given by Eq.~(\ref{Masses}) with the appropriate derivatives of the $L$-dependent potentials $U_\Lambda^L(\rho)$. All the dependencies 
on $q$ dropped out after performing the momentum trace.  
We calculate the free energy $\sigma(L)$ given by Eq.~(\ref{sigma}) supplemented by the flow Eqs~(\ref{U_Casi1}) and (\ref{U_Casi2}), which are integrated for a sequence of $L$-values. 
Eq.~(\ref{sigma}) indicates that the thermal energy $k_BT$ sets the scale for $\sigma(L)$ and therefore also for the Casimir force. We plot the 
difference $\left[U_{\Lambda\to 0}^L(0)-U_{\Lambda\to 0}^{\infty}(0)\right]$ versus $L$ in the double-logarithmic scale, as exemplified in Fig.~4 for $d=2$ and $N=1$. 
The points tend to lie on a straight line for 
sufficiently large $L$, indicating scaling behavior.
\begin{figure}[h]
\includegraphics[width=8.5cm]{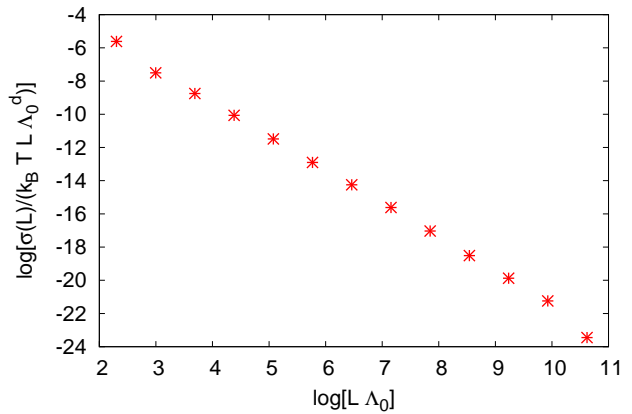}
\caption{Logarithmic plot of the difference $\left[U_{\Lambda\to 0}^L(0)-U_{\Lambda\to 0}^{\infty}(0)\right]$ versus $L$ for $d=2$, $N=1$. Fitting a straight line to the scaling region yields 
the slope 
$-a=-2.0011\pm 0.0092$ and the free coefficient $b=-1.323\pm0.045$. This translates to the computed Casimir force decaying as $L^{-a}$ and the Casimir amplitude $\Delta_f(2,1)=0.266\pm 0.012$, 
which compares well to the exact result $\Delta_f^{exact}=\frac{\pi}{12}\approx 0.262$.\cite{Cardy_86} Deviation from scaling behavior occurs for large and small $L$, where the condition $\xi\gg L\gg \Lambda_0^{-1}$ 
becomes violated.}
\end{figure} 
By fitting a straight line in the regime $\xi\gg L\gg \Lambda_0^{-1}$, we find the slope $-a=-2.0011\pm 0.0092\approx -2$ and the free coefficient $b=-1.323\pm0.045$. This translates to the computed Casimir force decaying as 
$L^{-a}$. The critical Casimir force amplitude $\Delta_f$ is defined via 
\begin{equation}
 f_C(L)=-\frac{\partial\sigma}{\partial L}\sim-k_BT\Delta_f L^{-a} \textrm{\hspace{1mm} for large \hspace{0.5mm}} L\;, 
 \label{F_C}
\end{equation}
and the general expectation is that $a=d$. The Casimir amplitude can be computed from the fit coefficient $b$. In the present case $d=2$, $N=1$ we find $\Delta_f(2,1)=0.266\pm 0.012$, 
which compares well to the exact result $\Delta_f^{exact}=\frac{\pi}{12}\approx 0.262$.\cite{Cardy_86} 

The above discussion exemplifies the procedure for the case $d=2$, $N=1$, where all results can be compared to exact values. 
In principle, the present approximation is least reliable for $d=2$, but nonetheless our result turns out 
to agree with the exact one. The method is obviously not restricted to this case and below we present our results 
for the Casimir force amplitudes $\Delta_f(d,N)$ for $N=1$ and $N=2$ and varying dimensionality $d$ between $2$ and $3$. For each considered $d$ we plot the difference 
$\left[U_{\Lambda\to 0}^L(0)-U_{\Lambda\to 0}^{\infty}(0)\right]$ vs $L$ (in a logarithmic scale). By fitting a straight line we find the free coefficient $b$ together with the error and use 
Eq.~(\ref{F_C}) to extract the Casimir amplitude supplemented with the corresponding error.

\subsection{Ising universality class} 
 In this subsection we discuss the Ising universality class fixing $N=1$, and perform the procedure outlined above varying $d$ between 2 and 3. For these values of $d$, the obtained 
 values of $a$ which govern the decay of the Casimir force agree with the expected behavior $a=d$. The Casimir force amplitude as a function of $d$ is plotted in Fig.~5. 
\begin{figure}[h]
\includegraphics[width=8.5cm]{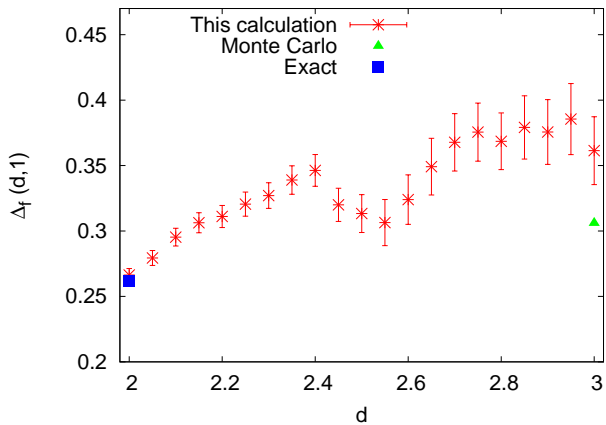}
\caption{The Casimir force amplitude $\Delta_f(d,1)$ as function of dimensionality $d$ in the Ising universality class $N=1$. The extra point at $d=2$ (square) corresponds to the exact result, 
and in $d=3$ (triangle) to the 
prediction of MC simulations of Ref.~\onlinecite{Vasilyev_09}. }
\end{figure} 
As already mentioned, the computed value of $\Delta_f$ in $d=2$ agrees with the exact result $\Delta_f^{exact}=\frac{\pi}{12}$.\cite{Cardy_86} For $d=3$ we obtain $\Delta_f(3,1)=0.361\pm 0.026$. 
This may be compared to 
numerical MC simulations of Ref.~\onlinecite{Vasilyev_09}, which give $\Delta_f\approx 0.306$.

The variation of $\Delta_f(d,1)$ is much 
more pronounced within the range $d\in [2,2.6]$  than for $d>2.6$, where a constant value is compatible with our results within the errorbars. The relative error resulting from the fitting procedure outlined in the previous 
section is about $4\%$ around $d=2$ and increases to $7\%$ around $d=3$. The accuracy decreases for growing $d$ because the quantity $\sigma (L)$ computed via Eq.~(\ref{sigma}) becomes smaller 
(at fixed $L$) and therefore the 
maximal value of $L$ that can be reliably handled numerically is lowered. 

Let us also compare the present result in $d=3$ with previous RG studies relying on the $\epsilon$-expansion. The two-loop result of Ref.~\onlinecite{Krech_92} retains terms to order $\epsilon$ and yields 
$\Delta_f(3,1)\approx 0.22$.
As discussed in Ref.~\onlinecite{Diehl_06} (see also Ref.~\onlinecite{Gruneberg_08}), 
the $\epsilon$-expansion is however problematic in the present case 
of periodic boundary conditions, yielding at three-loop level terms nonanalytic in $\epsilon$. The next-to-leading contribution is of the order $\epsilon^{3/2}$ and leads to $\Delta_f(3,1)$ corrected to 
$0.39$ in 
$d=3$. 

\subsection{XY universality class} 
The computation proceeds along the same lines in the present case, where we put $N=2$. This yields the extra contribution to the effective potential flow described by Eqs~(\ref{U_Casi1}) and 
(\ref{U_Casi2}) resulting from the presence of the Goldstone mode. As already discussed, the approximation level is too low to access the Kosterlitz-Thouless physics in $d=2$ and we limit the range 
of $d$ to $d\in [2.2,3]$. The Casimir force amplitude $\Delta_f(d,2)$ is plotted as a function of $d$ in Fig.~6.
\begin{figure}[h]
\includegraphics[width=8.5cm]{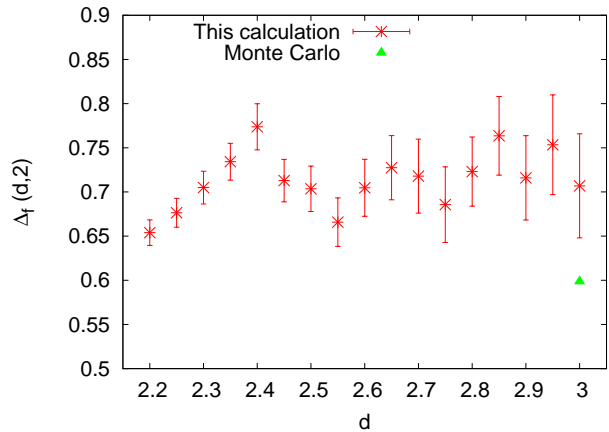}
\caption{The Casimir force amplitude $\Delta_f(d,2)$ as function of dimensionality $d$ in the XY universality class $N=2$. The extra point in $d=3$ (triangle) corresponds to the prediction of MC simulations. }
\end{figure} 
Similarly to the case $N=1$, we observe a variation in the dependence $\Delta_f(d,2)$ for $d< 2.6$, while for $d>2.6$ our results are consistent with $\Delta_f(d,2)=const$. We find 
$\Delta_f(3,2) = 0.707\pm 0.058$, which may be compared to the MC value\cite{Vasilyev_09} $\Delta_f^{MC}\approx 0.599$. Similarly to the case $N=1$, the MC result is lower and outside our errorbars. Our 
prediction is again closer to the MC results than the earlier perturbative RG studies, which yield\cite{Diehl_06, Gruneberg_08} $\Delta_f(3,2)\approx 0.84$.
The dependencies depicted in Figs 5, 6 are qualitatively similar, in particular we do not observe anything dramatic happening upon reducing $d$ towards $d=2$ in the $O(2)$ case (Fig.~6).

It is also worthwhile noting the relatively large errorbars resulting from the fitting procedure illustrated in detail at the beginning of the section. The errors account only for the numerical 
inaccuracies. We have nothing to say here about the errors caused by the approximation made while deriving the flow equations. As already remarked, the errorbars are wider for $d\approx 3 $ than in 
$d\approx 2$. The inaccuracies are larger than one might perhaps expect, because calculation of the Casimir amplitude requires finding the free coefficient in the fit straight 
line given only points far from $\ln(L\Lambda_0)=0$. For this reason a small tilt variation causes sizable changes in the point of interest, where the fit line intersects the $y$-axis. This results in significantly 
worse accuracy in determining the free coefficient than the slope.

\subsection{Role of the anomalous dimension} 
The calculation presented in Sec.~III, IVA and IVB relies on the ansatz (\ref{LPA_ansatz}), and, as described in Sec. III, gives inaccurate values of the bulk critical indices for $d\approx 2$. On the other 
hand, the value of the Casimir amplitude obtained for $d=2$, $N=1$ compares very well to the exact number. The quality of the result for $\Delta_f$ in this case does not therefore seem to suffer from the 
inaccuracies in resolving the bulk critical behavior. Here we present an extension of the ansatz (\ref{LPA_ansatz}) accounting for the flow of the $Z$-factor. This refinement leads to nonzero bulk $\eta$ 
exponent and much more accurate estimate of the correlation length exponent $\nu$. We perform an explicit computation for $d=2$ and $N=1$, where $\eta$ is largest, and show that the inclusion of $\eta$ has 
in fact no impact on our estimates of the Casimir forces. 

We now consider Eq.~(\ref{LPA_ansatz}) where the coupling $Z_0$ depends on the cutoff scale $\Lambda$, therefore making the replacement $Z_0\to Z_\Lambda^L$, where $Z_{\Lambda_0}^L=Z_0=1$. The flow 
equations for the effective potential (\ref{U_Casi1}) and (\ref{U_Casi2}) remain valid upon substituting $Z_0\to Z_\Lambda^\infty$ and $Z_0\to Z_\Lambda^L$, respectively. So does Eq.~(\ref{u_flow}), where 
the term $(d-2)$ becomes replaced by $(d+\eta-2)$, and $\eta$ is defined via $\eta = (\partial \ln(Z_\Lambda))/(\partial s)$.\cite{Berges_02} The flow of the $Z$-factor is extracted along the standard 
reasoning.\cite{Berges_02, fRG_reviews} 
Functional differentiation of Eq.~(\ref{Wetterich_eq}) yields the flow of the inverse propagator $\Gamma^{(2)}[\vec \phi]$, and the Laplacian of the resulting equation at 
$\rho=\rho_0$ corresponding to the minimum and at zero momentum determines $\partial_s Z_\Lambda^L$. The truncation level and the way of extracting $\partial_s Z_\Lambda^L$ are fully equivalent to the 
procedure 
of Refs.~(\onlinecite{Jakubczyk_09, Yamase_11}). For the flowing $Z$-factor at $L<\infty$ we obtain
\begin{eqnarray}
\partial_s Z_\Lambda^L = \frac{-A_{d-1}}{L}\frac{2\rho_0 Z_\Lambda^L\Lambda^2 [2\rho_0{U_{\Lambda}^L}'''(\rho_0)+3{U_{\Lambda}^L}''(\rho_0)]^2}
{[Z_\Lambda^L\Lambda^2+{U_{\Lambda}^L}'(\rho_0)+2\rho_{0} {U_{\Lambda}^L}''(\rho_0)]^5}\times \nonumber\\
\times \Big[-2Z_\Lambda^L\left(Z_\Lambda^L\Lambda^2+{U_{\Lambda}^L}'(\rho_0)+2\rho_{0} {U_{\Lambda}^L}''(\rho_0)\right)\Lambda^{d-1}\times \nonumber\\
\times \left(1+2\mathcal{S}(d)\right)+ \frac{8{Z_\Lambda^L}^2}{d+1}\Lambda^{d+1}\left(1+2\mathcal{S}(d+1)\right)\Big]\; . \;\;\;\;
\end{eqnarray}
We introduced 
\begin{equation}
\mathcal{S}(d) = \sum_{n=1}^{n_{max}}\left(1-(\frac{2\pi n}{\Lambda L})^2\right)^{(d-1)/2} \;, 
\end{equation}
where in turn $n_{max}=\frac{\Lambda L}{2\pi}$. The corresponding equation for infinite $L$ reads
\begin{eqnarray}
 \partial_s Z_\Lambda^\infty = \frac{-A_{d}(2\rho_0) Z_\Lambda^\infty\Lambda^2 [2\rho_0{U_{\Lambda}^\infty}'''(\rho_0)+3{U_{\Lambda}^\infty}''(\rho_0)]^2}
 {[Z_\Lambda^\infty\Lambda^2+{U_{\Lambda}^\infty}'(\rho_0)+2\rho_{0} {U_{\Lambda}^\infty}''(\rho_0)]^5} \times \nonumber \\
\times \Big[-2Z_\Lambda^{\infty}\Lambda^d [Z_\Lambda^\infty\Lambda^2+{U_{\Lambda}^\infty}'(\rho_0)+2\rho_{0} {U_{\Lambda}^\infty}''(\rho_0)] + \nonumber \\
\frac{8{Z_\Lambda^\infty}^2 \Lambda^{d+2}}{d+2}\Big] \;\;\;\;\;\;
\end{eqnarray} 
which is equivalent to an analogous expression in Ref.~\onlinecite{Jakubczyk_09}. The flowing bulk anomalous dimension now follows from $\partial_s \ln(Z_\Lambda^\infty)$. Upon tuning the system to bulk 
criticality by the dichotomy procedure of Sec.~II, the flowing $\eta$ attains a fixed-point value $\eta\approx 0.43$ (see Fig.~ 7). 
\begin{figure}[h]
\includegraphics[width=8.5cm]{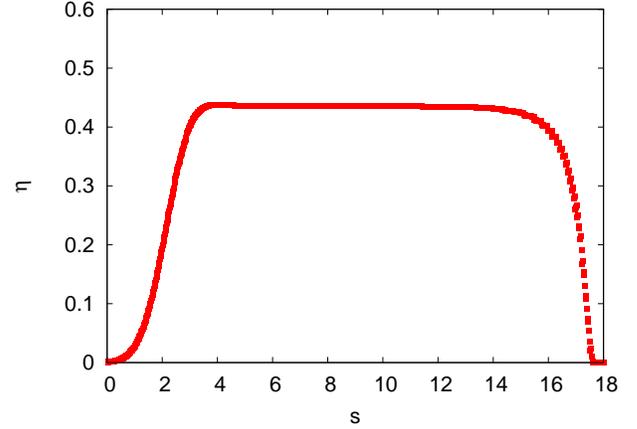}
\caption{The flowing anomalous dimension $\eta$ at criticality. The quantity $\eta (s)$ ultimately departs from the fixed point at $\eta\approx 0.43$ since the system is tuned to the critical manifold with 
a finite accuracy.}
\end{figure}

As expected,\cite{Berges_02} the 
present truncation overestimates the value of $\eta$. For the correlation length exponent we obtain $\nu\approx 0.96$, close to the correct value $1$. A similar calculation carried out in 
$d=3$,\cite{Jakubczyk_08} 
(applying a vertex expansion on top of the derivative expansion) yielded $\eta$ overestimated by a factor of nearly 2. 
The quality of the results for bulk critical exponents in $d=3$ is comparable to $\epsilon$ - expansion at second order, but the present approach performs much better than perturbative RG for $d=2$. 

We repeated the computation of the Casimir amplitude for the Ising universality class in $d=2$ within the present truncation capturing nonzero $\eta$. 
For the considered case $\eta$ is known to take the largest value and the local potential approximation is least reliable for resolving the bulk critical behavior. 
Nonetheless the obtained points in the scaling region lie practically on top of those plotted in Fig.~4 and calculated within the local potential approximation. 
We conclude, that including $\eta$ does not influence the asymptotic behavior of the Casimir forces.


\section{Summary and discussion}
In this work we have performed a renormalization-group study of Casimir forces occurring in the classical $d$-dimensional $O(N)$ models confined to a slab-like geometry and subject to periodic boundary conditions. 
In this case the system preserves translational symmetry and the Casimir forces arise solely due to fluctuations. These are long-ranged and universal at criticality. Our approach relies on a 
truncation of the one-particle-irreducible variant of functional RG at leading order in the so-called derivative expansion. We believe this approach complements the MC calculations in that the directly 
accessible system sizes are by far larger. On the other hand, it does not suffer from complications occurring in perturbative RG treatments and 
related to nonanalyticities in the $\epsilon$-expansion, revealed beyond two loops. It is also applicable both in $d$ close to 2 and 3, and we scanned a range of $d$-values in our calculations.  

Our results agree with the 
expectation that the critical Casimir force decays with the separation $L$ between the walls as $L^{-d}$ and the thermal energy $k_BT$ sets the relevant scale. We have computed the critical 
Casimir force amplitudes $\Delta_f(d,N)$ for 
the Ising 
($N=1$) and XY ($N=2$) universality classes varying dimensionality between 2 and 3 and remaining in the symmetric phase, but very close to the critical point. Our findings in $d=2$ 
compare well to the exact results.
In $d=3$ the computed Casimir amplitudes reasonably agree with numerical simulations, they are also significantly  
closer to MC than the earlier field-theoretic RG calculations.  The level 
of agreement (in particular in $d=2$ and $N=1$) suggests that the details of the momentum dependencies of the correlation function are not relevant for the magnitude of the Casimir forces at criticality. 
This may seem somewhat surprising, considering that the presently studied effect arises exclusively due to critical fluctuations.
Indeed, our 
calculation neglects the nonzero anomalous dimension $\eta$ whatsoever and also gives incorrect values of the correlation length exponent $\nu$ in $d=2$. Still, the results agree well with the exact 
findings in $d=2$. This result suggests considering the Casimir amplitudes among the critical properties which are insensitive to the presence of anomalous propagator scaling (nonzero $\eta$) 
quite alike, for example, the power laws describing the Casimir forces' decay or the universal shapes of the transition lines in quantum-critical systems.

To verify this, we performed an additional calculation in $d=2$, $N=1$ capturing the anomalous dimension. Althought this refinement leads to a much better resolution of the bulk critical behavior, it 
has practically no impact on our results referring to the Casimir forces. 

From a principal point of view, the present approach is applicable for all the $O(N)$ models at arbitrary dimensionality $d$ with periodic boundary conditions. 
One exception is the Kosterlitz-Thouless case, which requires going to higher order in derivative expansion. In practice however, the computation of the Casimir forces requires increasing numerical 
accuracy with growing $d$, and may become problematic already at $d=4$. Another interesting regime is $N\to\infty$, where one may again compare to exact results. 


It might be interesting to extend the present work to account for boundary conditions explicitly breaking translational symmetry. This problem is of interest also from the point of view of comparison to 
experiments performed on helium and fluid mixtures. The other interesting extensions which should be within the range of the present approach include computations within the ordered phase, where the soft 
fluctuations arise due to continuous symmetry breaking, and extracting the scaling functions for the Casimir force.

\begin{acknowledgments}
We thank W. Metzner and P. Nowakowski for very useful discussions and acknowledge funding by the National Science Centre via 2011/03/B/ST3/02638.
\end{acknowledgments}

\end{document}